\documentclass[pre,aps]{revtex4}
\usepackage{graphicx}

\begin{document}

\title{Molecular dynamics simulation of chains mobility in polyethylene crystal}

\author{V.\,I.\,Sultanov}
\affiliation{Science for Technology LLC, Leninskiy pr-t 95, 119313
Moscow, Russia}
\author{V.\,V.\,Atrazhev}
\email{vvatrazhev@deom.chph.ras.ru}
\affiliation{Institute of
Biochemical Physics of RAS, Kosygin str.4, 119991, Moscow,
Russia.}
\affiliation{Science for Technology LLC, Leninskiy pr-t
95, 119313 Moscow, Russia}
\author{D.\,V.\,Dmitriev}
\affiliation{Institute of Biochemical Physics of RAS, Kosygin
str.4, 119991, Moscow, Russia.}
\affiliation{Science for
Technology LLC, Leninskiy pr-t 95, 119313 Moscow, Russia}
\author{S.\,F.\,Burlatsky}
\affiliation{United Technologies Research Center, 411 Silver Lane,
East Hartford, CT 06108, USA}
\date{}

\begin{abstract}
The mobility of polymer chains in perfect polyethylene (PE)
crystal was calculated as a function of temperature and chain
length through Molecular dynamics (MD) in united atom
approximation. The results demonstrate that the chain mobility
drastically increases in the vicinity of the phase transition from
the orthorhombic to quasi-hexagonal phase. In the quasi-hexagonal
phase, the chain mobility is almost independent on temperature and
inversely proportional to the chain length.
\end{abstract}

\maketitle

Mechanical properties of semi-crystalline polymers below the
melting temperature are strongly governed by morphology of
crystallites that depends on how the melt was prepared and
treated. Polymer crystallites can undergo a variety of structural
phase transitions that is a subject of extensive studies. PE is
widely used as a `model system' for high-crystallinity polymer
\cite{Zubova12}. At normal conditions (room temperature and
atmospheric pressure), the PE crystal is in orthorhombic
\cite{Bruno} or monoclinic phase \cite{Yemni}. The PE crystals
undergo phase transition to quasi-hexagonal phase at elevated
pressure (more than 400 MPa) and temperature ($>520$K)
\cite{Bassett,Hikosaka,Langen}.

Quasi-hexagonal phase has been found in numerous diverse polymeric
systems \cite{Ungar}. One common property of the quasi-hexagonal
polymeric phase is some degree of conformational disorder, either
in the main chain or in the side groups or in both. Compared with
the ordered crystalline state, there is a high degree of molecular
mobility, with the chain performing both rotational and
translational motion. Translational chain mobility enables easy
formation of extended chain crystals in polymers that exhibit the
quasi-hexagonal phase; isothermal extension of the initially
folded chains has been shown to occur in the quasi-hexagonal phase
\cite{Ungar}. The PE chains in both monoclinic and quasi-hexagonal
crystals are parallel to each other. However, the monomers of the
chain in monoclinic phase predominantly belong to the same plane,
while the monomers of the chain in quasi-hexagonal phase are
randomly oriented. Therefore, the chains in quasi-hexagonal phase
form close packing of rods \cite{Ungar}.

Diffusion rate of PE chains in quasi-hexagonal phase was measured
experimentally in Ref.\cite{Langen} using proton spin-lattice
relaxation experiments and significant increase in the rate of
chain diffusion from about $10^{-12} cm^2/s$ (orthorhombic phase)
to $10^{-9} cm^2/s$ (quasi-hexagonal phase) was observed. The
molecular dynamics modeling and understanding of the atomistic
mechanisms of high chains mobility in the quasi-hexagonal phase is
the purpose of the current work.

The LAMMPS software package \cite{Plimpton} was utilized for
molecular dynamics simulations. Polyethylene chains were modeled
in united atom version of Dreiding forcefield \cite{Mayo}. The
Nose-Hoover style thermostat and barostat were used in these
calculations \cite{Shinoda}. The modeled samples were comprised of
64 polyethylene chains with 36, 100 and 200 carbon atoms in each
chain. The simulations were performed in periodic boundary
conditions in all directions. The chains were made `infinite' via
binding the last carbon atom of each chain with the first carbon
atom of the closest image of the same chain in c direction. We
used orthorhombic phase of PE crystal as an initial state of our
simulations. The front and side views of the initial configuration
of the modeled sample with 100 carbon atoms in each chain are
shown in Figure 1(a). After initial geometrical optimization of
the system, the MD trajectory was run in NPT ensemble. We use
below the relative temperature, $\tau$, normalized by the
Lennard-Jones interaction constant, $\varepsilon$;
$\tau=T/\varepsilon$.

\begin{figure}[tbp]
\includegraphics[width=3in,angle=0]{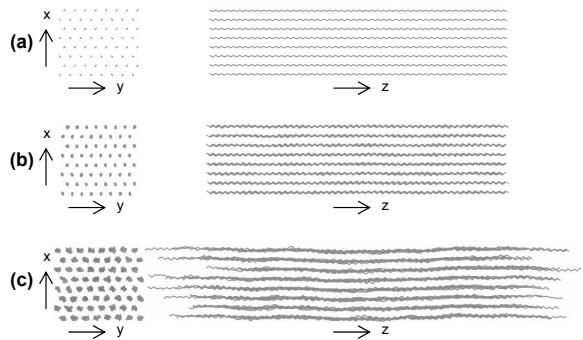}
\caption{Crystalline sample: (a) orthorhombic phase, $\tau = 0$;
(b) monoclinic phase, $\tau = 3$, $t=1$ ns, (c) quasi-hexagonal
phase, $\tau = 5$, $t = 1$ ns.} \label{Fig1}
\end{figure}

Although we started from the orthorhombic phase, the transition
into the monoclinic phase occurred within 50 ps from start of MD
simulation at the temperatures below $\tau=4.5$. The snapshot
(front and side view) of the sample configuration at temperature
$\tau = 3$ and time $t = 1$ ns after initial time moment is shown
in Figure 1(b). The monomers of individual chains predominantly
belong to one plane. Thermal motion results in fluctuations of the
monomer near the chain plane, as seen at the front view of Figure
1(b), and in the transverse waves that are observed at the side
view Figure 1(b). The transition into the quasi-hexagonal phase
was observed at the temperature above $\tau=4.5$. The snapshot
(front and side view) of the sample configuration at temperature
$\tau=5$ and time $t = 1$ ns after initial time moment is shown in
Figure 1(c). The monomers in quasi-hexagonal phase are randomly
oriented although the chains in the crystal are parallel to each
other on the average. The transverse waves are also observed in
quasi-hexagonal phase at the side view of Figure 1(c).

The ratio of the elementary crystal cell dimensions transversal to
the chain direction, $b/a$, is equal to $\sqrt{3}\approx1.732$ in
the quasi-hexagonal phase. Calculated $b/a$ averaged through MD
trajectory in NPT ensemble as a function of temperature is
presented in Figure 2. The transition from monoclinic to
quasi-hexagonal phase occurs in the temperature range from
$\tau=4$ to $\tau=4.5$ Lennard-Jones units as seen in Figure 2.
The transition manifests itself in the abrupt change of the system
sizes in transversal directions, which indicates the change of the
chain packing in the crystal. The ratio $b/a=1.732$ stated above
the transition temperature implies the quasi-hexagonal packing of
the chains. Qualitatively similar dependence of $b/a$ on
temperature for PE crystal were obtained in united atom molecular
dynamic simulation in \cite{Zubova08} (see Fig. 1 therein). Minor
quantitative differences between Figure 2 of this paper and Figure
1 in \cite{Zubova08} are caused by the difference in force fields
used in this work and in \cite{Zubova08}.

\begin{figure}[tbp]
\includegraphics[width=3in,angle=0]{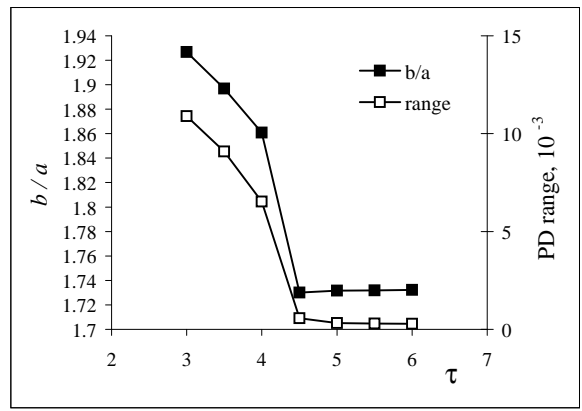}
\caption{Temperature dependencies of the elementary cell
dimensions ratio $b/a$ and the probability density range of
setting angle.} \label{Fig2}
\end{figure}

To study the orientation order in the phases the distribution of
the setting angle was calculated following \cite{Zubova08}. The
distribution of the setting angle is a sum of $\delta$-functions
in totally orientation ordered phase and is a constant in
disordered phase. The local setting angle $\alpha$ is defined for
each atom as follows. For each bond angle formed by three
neighboring atoms in one chain, a bisector vector is constructed,
which is multiplied by $(-1)^n$ ($n$ is the number of an atom in
the chain). The angle $\alpha$ is the angle between the x-axis and
the projection of this vector onto $xy$ plane. The distributions
of setting angle for three temperatures are shown in Figure 3.
Qualitatively,  Figure 3 looks similar to Figure 2 in
\cite{Zubova08}. The quantitative differences should be attributed
to the difference of the employed force fields.

\begin{figure}[tbp]
\includegraphics[width=3in,angle=0]{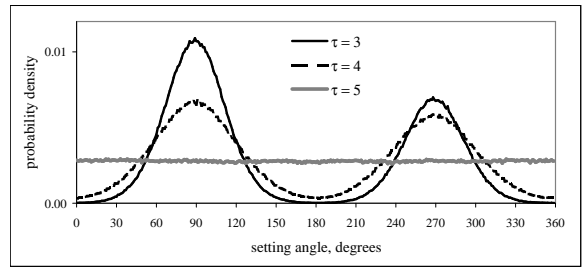}
\caption{Local setting angle distribution function for three
temperatures.} \label{Fig3}
\end{figure}

In monoclinic phase ($\tau=3$ and $\tau=4$), the setting angle is
distributed near $\alpha=0^\circ$ and $\alpha=180^\circ$ and the
width of distribution decreases with decrease of temperature. In
quasi-hexagonal phase ($\tau=5$), the distribution of setting
angle is uniform that indicates the loss of orientation order. For
quantitative characterization of uniformity of the setting angle
distribution, we utilized the range of distribution. The range is
equal to the maximal value of the setting angle minus the minimum
value of the setting angle. For totally inform distribution the
range is equal to zero. The range of the setting angle
distribution as a function of temperature is plotted in Figure 2.
The range tends to zero simultaneously with tending of $b/a$
to $1.732$, i.e. when the crystal undergoes the phase transition
to the quasi-hexagonal phase. That indicates the vanishing of the
orientation order of monomers at the phase transition to the
quasi-hexagonal phase.

The relatively large deviation of the chains from their initial
positions (Figure 1(a)) in axial direction is observed in
quasi-hexagonal phase (Figure 1(c)) while the chains keep their
initial positions in monoclinic phase (Figure 1(b)). That
indicates the increased chains mobility in axial direction in the
quasi-hexagonal phase. For quantitative characterization of the
chains mobility, we calculated the diffusion coefficient of the
chain in axial direction as a function of temperature. Diffusion
coefficient was calculated from MD trajectories as follows. To
track the large-scale displacement of the center of mass of the
chain we track the displacement of one selected atom of this
chain. The axial coordinates, $z_j(t_i)$, of selected atom of each
chain were recorded at each MD time step, $t_i$. The mean square
displacement of the chains as a function of time was calculated as
\begin{equation}
\langle\Delta z^2(t_i)\rangle = \sum_{j=1}^{N_c} (z_j(t_i)-z_j(0))^2
\end{equation}
where $N_c$ is the number of chains in the system.

The dependencies $\langle\Delta z^2(t)\rangle$ for three
temperatures are presented in Figure 4 for the crystal with 100
carbon atoms in each chain.

\begin{figure}[tbp]
\includegraphics[width=3in,angle=0]{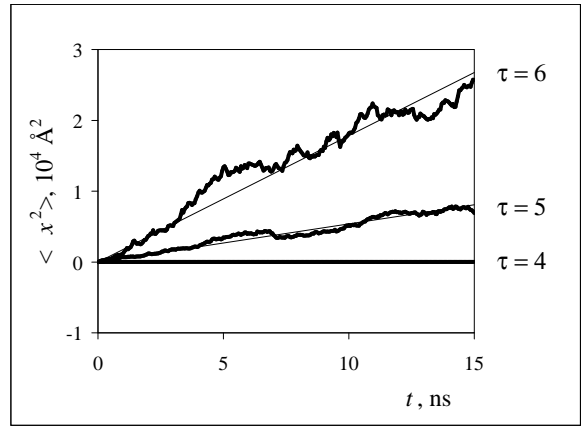}
\caption{Calculated from MD trajectories mean square displacements
of the chains, $\langle\Delta z^2(t)\rangle$, as functions of time
for several temperatures. The number of carbon atoms in the chain
is 36.} \label{Fig4}
\end{figure}

The mean square displacements in Figure 4 were approximated by
linear dependencies on time, $\langle\Delta z^2(t)\rangle=2Dt$.
Diffusion coefficient of the chains, $D$, is calculated from the
slope of the linear trend of $\langle\Delta z^2(t)\rangle$.
Calculated dependencies of $D$ on the temperature for chain with
100 carbon atoms is plotted in Figure 5. The range of setting angle
distribution as a function of temperature is also shown in the
same plot. The diffusion coefficient, $D$, abruptly increases in
narrow region in the vicinity of the phase transition temperature
from monoclinic to quasi-hexagonal phase as seen from Figure 5.
The chain mobility at the temperature $\tau=6$ (quasi-hexagonal
phase) is almost four orders of magnitude higher than that at the
temperature $\tau=4$. Diffusion coefficient, $D$, is plotted with
the setting angle distribution range in Figure 5 to emphasize the
correlation between the orientation order and the chain mobility
in the crystal. Thus, the MD simulation predicts that the chain
mobility in PE crystal increases sharply with the vanishing of the
orientation order in the crystal at the transition to the
quasi-hexagonal phase.

\begin{figure}[tbp]
\includegraphics[width=3in,angle=0]{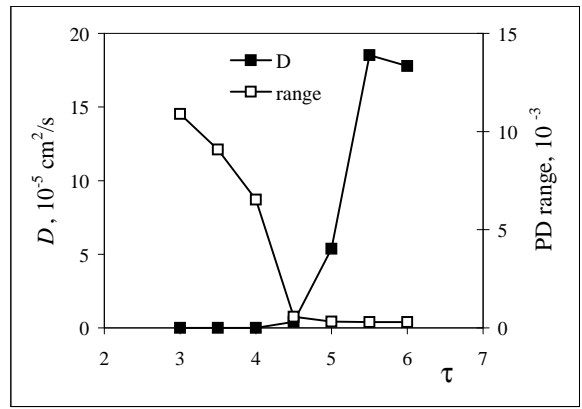}
\caption{Diffusion coefficient of the chains as a function of
temperature.} \label{Fig5}
\end{figure}

The chain mobility in the quasi-hexagonal phase at $\tau=6$ was
calculated for four chain lengths and plotted as a function of the
chain length in Figure 6. The chain mobility in quasi-hexagonal
phase is approximately inversely proportional to the chain length
as seen in Figure 6. That indicates that the chains move in
perfect quasi-hexagonal crystal as a whole without activation
barriers. The PE chains diffusion in the crystallites in
orthorhombic/monoclinic phase via the local topological defects
was studied in Refs.\cite{Zubova99,Zubova07} by molecular dynamics
simulations. The diffusion of the defects along the chain results
in chain longitudinal motion in the crystallite, that is similar
to reptation in polymer melt. The mechanism of the chain motion in
the quasi-hexagonal phase differs from that in orthorhombic or
monoclinic phases and does not involve the topological defects.
The rod-like close packing of chains in the quasi-hexagonal phase
indicates that the nearest chains are not sensitive to the details
of the inter-chain interaction. That implies the `sliding'
mechanism of the chains motion in quasi-hexagonal phase, when the
chains slide relative to each other without activation barriers.
The random collisions of the atoms of the chain with the atoms of
the nearest chains result in viscous friction. The friction force
is proportional to the chain length.

\begin{figure}[tbp]
\includegraphics[width=3in,angle=0]{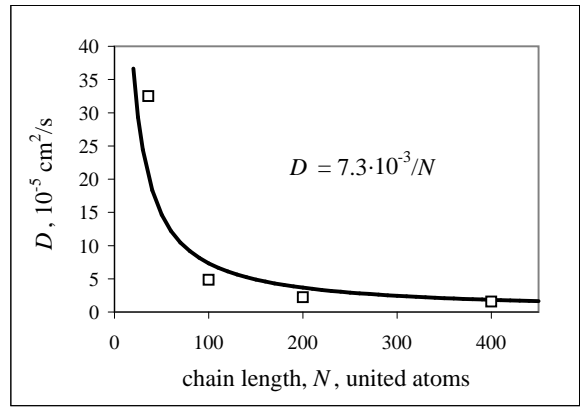}
\caption{Dependence of diffusion coefficient of the chains on
chain length at temperature $\tau = 5$ (hexagonal phase). Solid
line is a fit by dependence $D = const/N$.} \label{Fig6}
\end{figure}

In conclusion, our MD study shows that the mobility of PE chains in
the crystal increases sharply (by four orders of magnitude) at the
transition of the PE crystal to the quasi-hexagonal phase, where
the orientation order vanishes. The obtained results are in agreement
with the experimental observation \cite{Langen}.
The sliding of polymer chains through the crystallites plays a key
role in the so-called $\alpha_c$ structural transition in
semicrystalline polymers with high degree of crystallinity,
which is observed in dielectric relaxation measurements \cite{Boyd}.
Besides, the understanding of the mechanism of chain mobility in
crystallites can help to develop the microscopic theory
of semicrystalline polymers in high-elastic state \cite{we}.

The authors gratefully acknowledge Dr. M. McQuade and Dr. D.
Parekh of United Technologies Corporation for the interest to the
work, inspiring discussion and support, and Professor J. M. Deutch
of Massachusetts Institute of Technology and Dr. Charles Watson of
Pratt and Whitney for interesting discussion of the results, and the
Joint Supercomputer Center of Russian Academy of Sciences for the
computational resources granted.

\end{document}